\documentclass[preprint,12pt]{elsarticle}

\usepackage{amssymb}
\usepackage{lineno}

\journal{NIM B}

\begin{document}

\begin{frontmatter}



\title{Employing $p+^{58}$Ni elastic scattering for determination of $K$-shell ionization cross section of $^{58}$Ni$^{19+}$ in collisions with hydrogen gas target at 95 MeV/u}


\author[label0,label1,label2]{J. T. Zhang}
\author[label2]{K. Yue}
\author[label2]{C. J. Shao}
\author[label2,label7,label3]{X. L. Tu\corref{cor2}}
\cortext[cor2]{Corresponding author.}
\ead{tuxiaolin@impcas.ac.cn}
\author[label2]{Y. Y. Wang}
\author[label2]{P. Ma}
\author[label4,label2]{B. Mei}
\author[label2]{X. C. Chen}
\author[label2]{Y. Y. Yang}
\author[label2]{Z. Y. Sun}
\author[label2,label0]{M. Wang}
\author[label5]{V. P. Shevelko}
\author[label5]{I. Yu. Tolstikhina}
\author[label6]{Yu.  A. Litvinov}
\author[label2]{Y. H. Zhang}
\author[label2]{X. H. Zhou}

\address[label0]{Joint Department for Nuclear Physics, Lanzhou University and Institute of Modern Physics, Chinese Academy of Sciences, Lanzhou 730000, China}
\address[label1]{School of Nuclear Science and Technology, Lanzhou University, Lanzhou 730000, China}
\address[label2]{Institute of Modern Physics, Chinese Academy of Sciences, Lanzhou 730000, China}
\address[label7]{School of Nuclear Science and Technology, University of Chinese Academy of Sciences, Beijing 100049, China}
\address[label3]{Max-Planck-Institut  f\"ur Kernphysik, Saupfercheckweg 1, 69117 Heidelberg, Germany}
\address[label4]{Sino-French Institute of Nuclear Engineering and Technology, Sun Yat-sen University, Zhuhai 519082, China}
\address[label5]{P. N. Lebedev Physical Institute, Leninskii pr. 53, 119991 Moscow, Russia}
\address[label6]{GSI Helmholtzzentrum f\"ur Schwerionenforschung, D-64291 Darmstadt, Germany}

\begin{abstract}
We present a new experimental method for measuring inner-shell ionization cross sections of low-charged ions colliding with hydrogen gas target in a storage ring. The method is based on a calibration by the well-known differential cross sections of proton elastic scattering on nuclei. $K$-shell ionization cross section of 1047(100) barn for the 95 MeV/u $^{58}$Ni$^{19+}$ ions colliding with hydrogen atoms was obtained in this work. Compared to the measured ionization cross section, a good agreement is achieved with the prediction by the Relativistic Ionization CODE Modified program (RICODE-M).

\end{abstract}


\end{frontmatter}

% \linenumbers

\section{Introduction}
Inner-shell ionization cross sections for low-charged ions colliding with atoms are of great importance for understanding atomic structure and reaction mechanism~\cite{Garcia73}. They have already found application in estimation of beam loss resulting from interactions with residual gas in an accelerator~\cite{Shevelko18,Kaganovich05,Weber09,DuBois07}. Hydrogen gas (H$_2$) is one of major compositions of the residual gas in accelerator vacuum systems~\cite{Weber09}. Thus, the inner-shell ionization cross sections of the low-charged ions colliding with hydrogen atoms need to be investigated~\cite{Shevelko18,Kaganovich05,Weber09,DuBois07,Shevelko11,Shevelko01,Tolstikhina14}. It is essential for designing and operating next-generation accelerator facilities, such as the heavy-ion synchrotron SIS100 at the Facility for Antiproton and Ion Research (FAIR)~\cite{Henning08} and the accumulation and booster ring ABR45 at the High Intensity heavy ion Accelerator Facility (HIAF)~\cite{Yang13}. It is foreseen to employ low-charged ions to reduce space-charge effects to achieve the highest intensity primary beams.

Compared to proton impact~\cite{Lapicki08}, the ionization of the low-charged ions by hydrogen atom impact is different, since the influence of the electron in the hydrogen atom can not be ignored~\cite{Shevelko11}. As a result, accurate theoretical calculations for the inner-shell ionization cross sections in the ion-H collisions are more complex. To test theories, the experimental inner-shell ionization cross sections for the low-charged ions colliding with hydrogen atoms are required.
Heavy-ion synchrotrons can accelerate low-charged ions of interest to required energies. The ions are stored in ultra-high vacuum environment, thus preserving their charge states for sufficiently long time to perform experimental studies.
The Heavy-ion experimental storage rings equipped with internal gas-jet targets, such as ESR at GSI and CSRe at IMP, have been used to study ion-atom collisions~\cite{Weber09,DuBois07,Shao17,Eichler07}.
However, it is difficult to precisely determine target density, beam intensity and the overlap between the target and beam in the ion-atom collision experiments at the storage ring. Consequently, only a precision of about 30$\%$ could be reached for absolute ionization cross section measurements~\cite{Weber09,Eichler07}. This experimental precision is lower than a precision of a few percent for theoretical predictions on atomic cross sections~\cite{Glorius19,Herdrich17}.
The experimental challenge was also noted in the review paper~\cite{Eichler07}.

Various reaction channels may exist in the ion-atom collisions. One can calibrate unknown reaction cross sections by using the reaction channels with well-known cross sections.
Compared to strong interaction, electromagnetic interaction is well studied. Thus, theoretical atomic cross sections are often used to normalize less-accurately known nuclear reaction cross sections in nuclear reaction experiments~\cite{Glorius19,Mei15}. Many in-ring nuclear reaction experiments have been performed at the ESR ring~\cite{Glorius19,Mei15,Zamora17,Schmid15,Zamora16}.
In a pilot in-ring  $^{58}$Ni($p$, $p$) $^{58}$Ni elastic scattering experiment at the Cooler Storage Ring at the Heavy Ion Research Facility in Lanzhou (HIRFL-CSR)~\cite{Yue19}, in order to show a potential for extracting optical model potentials (OMPs) of unstable nuclei, a theoretical $K$-shell ionization cross section was applied to determine the reaction luminosity. A good agreement was achieved with the calculated differential cross sections of proton elastic scattering, which were obtained by using the global phenomenological OMP~\cite{Koning03}. Inspired by the work~\cite{Yue19}, a simple experimental method to determine the $K$-shell ionization cross sections of the low-charged ions with hydrogen atoms collisions is described in the present paper. Consequently, the precision of the inner-shell ionization cross section measurements for the ion-H collisions could be improved by a factor of about 3.

\section{Luminosity determination and discussion}
\begin{figure}[h]
\begin{center}
\includegraphics*[width=0.7\textwidth]{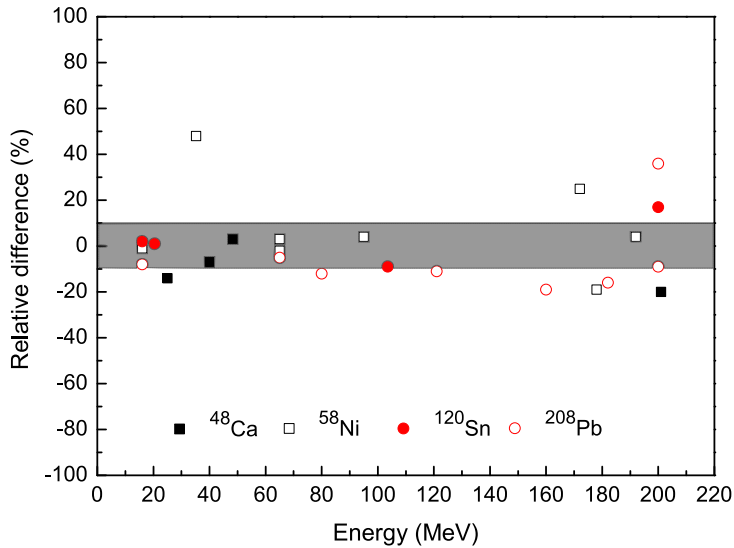}
\caption{Relative differences between experimental cross sections and the predicted cross sections by the KD03-OMP at small scattering angles for elastic proton scattering on $^{48}$Ca, $^{58}$Ni, $^{120}$Sn, and $^{208}$Pb with different impact energies. The gray area shows an uncertainty of 10$\%$. The experimental cross sections, published after 1980, are from CSISRC~\cite{CS}.}
\label{fig3}
\end{center}
\end{figure}

Elastic scattering on stable nuclei was investigated for about a century. 
The global phenomenological OMP parameters for proton elastic scattering have been extracted from experimental data with incident energies from 1 keV up to 200 MeV~\cite{Koning03,Li08}.
By using the global phenomenological OMP  parameters suggested by A. J. Koning and J. P. Delaroche (KD03)~\cite{Koning03}, reaction cross sections for proton scattering on stable and (near-) spherical nuclei were reproduced with a precision of 5$\%$ to 10$\%$ ~\cite{Koning03}. The recoiling protons at small scattering angles in the center-of-mass frame mainly originate from the well-known Rutherford scattering. Thus, there are no obvious differences for the differential cross sections of proton elastic scattering at small scattering angles between local, global and even different OMP model predictions~\cite{Koning03,Li08,Xu16}. The small scattering angle in this work is understood as the angle which is smaller than the angle at the first minimum inflection point of the differential cross section curve.
The cross sections of proton elastic scattering predicted by KD03-OMP at small scattering angles were compared to the corresponding experimental cross sections $\sum\sigma(\theta)$ for $^{48}$Ca, $^{58}$Ni, $^{120}$Sn, and $^{208}$Pb with different impact energies. Figure~\ref{fig3} shows, a precision of 10$\%$ could be achieved for the elastic scattering cross sections predicted by the KD03-OMP at small scattering angles in the center-of-mass frame.

%%%%%%%%%%%%%%%%%%%%%%%%%%%%%%%%%% fig4 %%%%%%%%%%%%%%%%%%%%%%%%%%%%%%%%%%%
%%%%%%%%%%%%%%%%%%%%%%%%%%%%%%%%%%%%%%%%%%%%%%%%%%%%%%%%%%%%%%%%%%%%%%%%%%%

The present method can be applied to extract the ionization cross sections for low-charged ions, especially for stable and (near-) spherical atomic nuclei.
In contrary to the method used in~\cite{Yue19}, the key point of this work is, that the luminosity needed to extract the ionization cross sections of the low-charged ions is precisely determined through the differential cross sections of proton elastic scattering on the corresponding low-charged ions. Consequently, parameters related to luminosity, such as the gas target density, the beam intensity and their special overlap, are no longer required to determine the inner-shell ionization cross sections in the internal gas-jet target experiments~\cite{Weber09,DuBois07,Eichler07}.

To illustrate the feasibility of this method, the experimental data from~\cite{Yue19} was reanalyzed.
Please note, that the differential cross sections used in this work are taken from the global phenomenological OMPs ~\cite{Koning03}. They are not the values obtained in~\cite{Yue19}. Thus, the results obtained in both works are independent from each other. 
The experiment in~\cite{Yue19} was conducted at the experimental storage ring (CSRe)~\cite{Xia02}. 
The $^{58}$Ni$^{19+}$ beam with energy of 95 MeV/u repeatedly interacted with the internal hydrogen molecular gas target in the CSRe. 
In the following we neglect the molecular bindings and consider the collisions to happen on H atoms.
$K$-shell x rays and proton recoils with small scattering angles from the $^{58}$Ni$^{19+}$-H collisions were measured simultaneously. More details on the experiment can be found in~\cite{Yue19}.

%%%%%%%%%%%%%%%%%%%%%%%%%%%%%%%%%% fig4 %%%%%%%%%%%%%%%%%%%%%%%%%%%%%%%%%%%
\begin{figure}[h]
\begin{center}
\includegraphics*[width=0.7\textwidth]{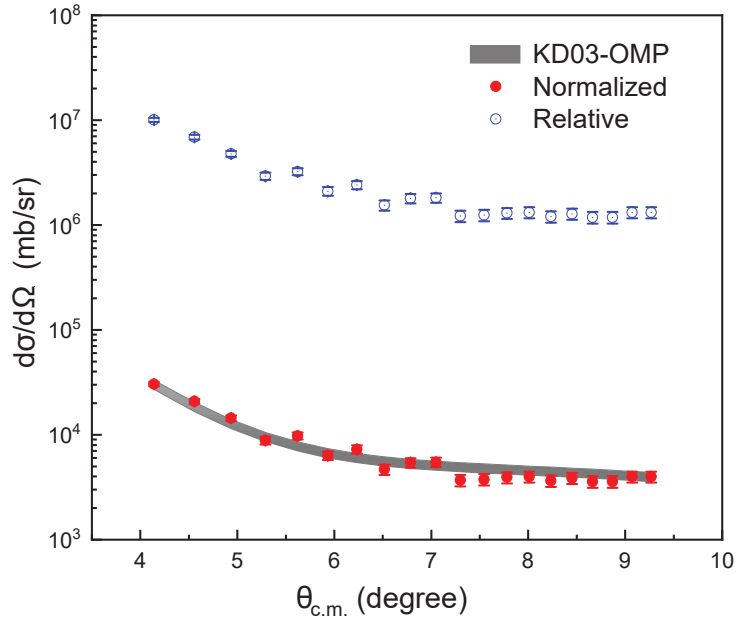}
\caption{Differential cross sections as a function of the proton scattering angle in the center-of-mass frame. The gray area shows the error band of 10$\%$ for the differential cross sections predicted by the KD03-OMP. The best agreement between normalized cross sections and the reference cross sections (see text) is achieved with an integrated luminosity of 3.31$\times$10$^{29}$ cm$^{-2}$. }
\label{fig4}
\end{center}
\end{figure}
%%%%%%%%%%%%%%%%%%%%%%%%%%%%%%%%%%%%%%%%%%%%%%%%%%%%%%%%%%%%%%%%%%%%%%%%%%%

%%%%%%%%%%%%%%%%%%%%%%%%%%%%%%%%%% tab1 %%%%%%%%%%%%%%%%%%%%%%%%%%%%%%%%%%%
\begin{table*}
\begin{center}
\caption{The OMP parameters used~\cite{Koning03} for the differential cross section calculation.}
\footnotesize
\renewcommand\tabcolsep{1 mm}
\begin{tabular}{cccccccccccc}
\hline\hline
~$V_V$&~$W_V$&~$r_V$&~$a_V$&~$W_D$&~$r_D$&~$a_D$&~$V_{SO}$&~$W_{SO}$&~$r_{SO}$&~$a_{SO}$&~$r_C$\\
~(MeV)&~(MeV)&~(fm)&~(fm)&~(MeV)&~(fm)&~(fm)&~(MeV)&~(MeV)&~(fm)&~(fm)&~(fm)\\
\hline
27.52&9.64&1.2&0.67&1.75&1.28&0.55&4.04&0.90&1.02&0.59&1.26\\

\hline\hline

\label{tab1}
\end{tabular}
\end{center}
\end{table*}
%%%%%%%%%%%%%%%%%%%%%%%%%%%%%%%%%%%%%%%%%%%%%%%%%%%%%%%%%%%%%%%%%%%%%%%%%%%

Table~\ref{tab1} lists the OMP parameters used in this work~\cite{Koning03}. The differential cross sections predicted by the KD03-OMP were obtained with the coupled-reaction channels program FRESCO \cite{Thompson}. These can also be calculated by using the nuclear reaction video (NRV) web calculator~\cite{NRV}. 
As Fig.~\ref{fig3} shows, the elastic scattering cross sections at small scattering angles predicted by the KD03-OMP have a precision of about 10$\%$.
Thus, they were chosen here as a reference for cross section normalization. The reaction luminosity can be determined by normalizing the relative experimental differential cross sections of proton elastic scattering to these reference cross sections. The relative experimental differential cross sections ($\frac{\Delta N_t}{\Delta t}$) can directly be extracted via the measured proton events, see~\cite{Yue19} and references cited therein.
To verify the effect of the background and the detection efficiency in the low energy region, the proton data were reanalyzed in this work. Only proton events within the energy peak, and with energies $T_k$ $\textgreater$ 453 keV were taken into account, see Fig.~2 in~\cite{Yue19}.  In this work, the relative experimental differential cross sections were normalized to the reference cross sections by adjusting the luminosity parameter to achieve a minimum $\chi^2$, 
%%%%%%%%%%%%%%%%%%%%%%%%%%%%%% equation2 %%%%%%%%%%%%%%%%%%%%%%%%%%%%%%%%%%
\begin{equation}
\chi^2=\sum_{i=1}^n\frac{[\frac{1}{L}(\frac{d\sigma}{d\Omega})^{r}_i-(\frac{d\sigma}{d\Omega})^{KD03}_i]^2}{[\frac{1}{L}\Delta(\frac{d\sigma}{d\Omega})^{r}_i ]^2}\quad,
\label{eq1}
\end{equation}
%%%%%%%%%%%%%%%%%%%%%%%%%%%%%%%%%%%%%%%%%%%%%%%%%%%%%%%%%%%%%%%%%%%%%%%%%%%
where $n$ is the number of data points taken for calculation of the differential cross sections, $L$ is the integrated luminosity,  ($\frac{d\sigma}{d\Omega})^{KD03}$ is the reference differential cross sections calculated by the KD03-OMP. ($\frac{d\sigma}{d\Omega})^{r}$ and $\Delta$($\frac{d\sigma}{d\Omega})^{r}$ are the relative experimental differential cross sections and the corresponding uncertainties, respectively.
The best agreement is achieved with an integration luminosity of 3.31$\times$ 10$^{29}$ cm$^{-2}$, see Fig.~\ref{fig4}. 

To extract the $K$-shell ionization cross section ($\sigma$), besides the number of the  measured $K$-shell x-rays, the fluorescence yield is also needed.
The $K$-shell fluorescence yield depends on the charge states of the ions~\cite{Doyle78}, which have been discussed in~\cite{Yue19} for the $^{58}$Ni$^{19+}$ ions. In this work, we calculated the $K$-shell fluorescence yields for the $^{58}$Ni$^{19+}$ ions by using the general-purpose relativistic atomic structure package (GRASP 2K)~\cite{Jonsson13}. A similar result as in~\cite{Yue19} was obtained. 
With the determined above luminosity, the $K$-shell ionization cross section of 1047(100) barn for the 95 MeV/u $^{58}$Ni$^{19+}$ ions colliding with hydrogen atoms was deduced.
The relative error of about 10$\%$ originates mainly from the uncertainty of the KD03-OMP calculations.
When ions loose/capture electrons, their trajectories are changed~\cite{Klepper92,Litvinov11}. In future, a position-sensitive particle detector~\cite{Najafi16} will be installed into the CSRe. Thus, the fully ionized ions will be measured directly behind the first dipole magnet downstream the gas target.

%%%%%%%%%%%%%%%%%%%%%%%%%%%%%%%%%% fig5 %%%%%%%%%%%%%%%%%%%%%%%%%%%%%%%%%%%
\begin{figure}[h]
\begin{center}
\includegraphics*[width=0.7\textwidth]{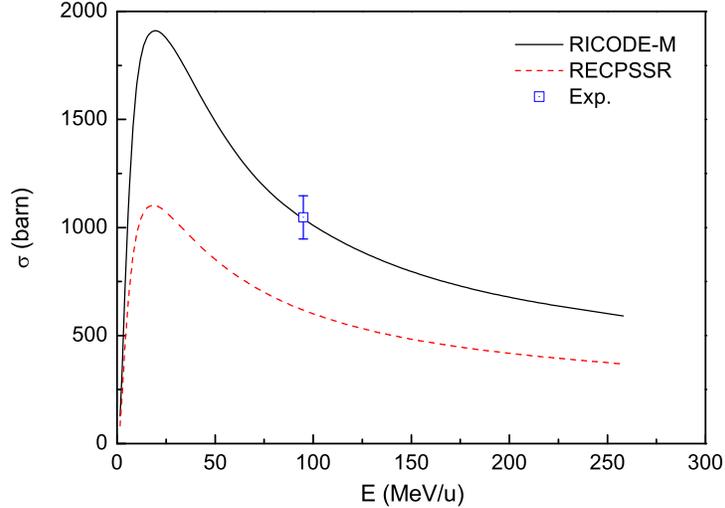}
\caption{Comparison of the experimental $K$-shell ionization cross sections (Exp.) for Ni$^{19+}$ ions induced by the impact on H atoms with the predictions of the RICODE-M theory~\cite{Tolstikhina14}. The cross sections calculated by the RECPSSR theory~\cite{Cipolla11} are for Ni atoms impacting on protons.}
\label{fig5}
\end{center}
\end{figure}
%%%%%%%%%%%%%%%%%%%%%%%%%%%%%%%%%%%%%%%%%%%%%%%%%%%%%%%%%%%%%%%%%%%%%%%%%%%

The obtained $K$-shell ionization cross section is compared to the prediction by the Relativistic Ionization CODE Modified program (RICODE-M)~\cite{Tolstikhina14} in Fig.~\ref{fig5}.
The RICODE-M~\cite{Tolstikhina14} is based on the relativistic Born approximation and particles are colliding with the relativistic wave functions. It is successfully used for calculations of electron-loss cross sections of heavy many-electron ions induced by neutral atoms in relativistic and non-relativistic energy domains.  For more details on the program the reader is referred to~\cite{Tolstikhina14}.
Figure~\ref{fig5} shows ionization cross sections of neutral Ni atoms induced by proton impact. They were calculated by the RECPSSR in the Inner-Shell Ionization Cross Sections (ISICS) program~\cite{Cipolla11}. The RECPSSR is based on the Energy-Loss Coulomb-Repulsion Perturbed-Stationary-State Relativistic (ECPSSR) theory where the relativistic projectile velocities are also considered~\cite{Lapicki08,Cipolla11}. The difference between two cross sections is related mainly to the difference in the effective charge of the H and proton targets. In addition, the ionization by the electron in the H atom also contributes to the cross section. The latter effect is known as "antiscreening" (see~\cite{Shevelko01} for more details).
The RICODE-M~\cite{Tolstikhina14} describes well the experimental cross section, see Fig.~\ref{fig5}.
The method presented here can be applied to measure not only the inner-shell ionization cross sections but also the radioactive electron capture cross sections for the ion-H collisions at low and intermediate energies.
The most important conclusion is, that the ionization cross sections are the same for nuclei with the same number of protons. However, the nuclear reaction cross sections, e.g., the differential cross sections of proton elastic scattering depend strongly on the proton-neutron asymmetry. Thus, the ionization cross sections known with high precision are important for in-ring nuclear reaction experiments to normalize nuclear reaction cross sections induced by light ions~\cite{Yue19}. Such elastic scattering reactions have already been used to study nuclear matter distribution~\cite{Zamora17,Schmid15,Sakaguchi17} and will continue in the future addressing, e.g., neutron skin thicknesses of heavy nuclei, that can be related to radius of neutron star~\cite{Fattoyev18}.

\section{Summary and outlook}
We presented a novel experimental method to measure the inner-shell ionization cross sections of $^{58}$Ni$^{19+}$ ions colliding with hydrogen gas target. The $K$-shell ionization cross section was calibrated by the well-known differential cross sections of proton elastic scattering on $^{58}$Ni nuclei. With this method, the inner-shell ionization cross sections for the ion-H collisions can routinely be determined with a precision of about 10$\%$. The $K$-shell ionization cross section of 1047(100) barn for the 95 MeV/u $^{58}$Ni$^{19+}$ ions colliding with hydrogen atoms was obtained. Compared to the  RICODE-M prediction, a good agreement has been reached.

This work is supported in part by the NSFC (11775273, U1932140), 
by the CAS Pioneer Hundred Talents Program, 
by the CAS Open Research Project of large research infrastructures, 
by the CAS Maintenance and Reform of large Research infrastructures (DSS-WXGZ-2018-0002), 
by the Max-Plank-Society, the Major State Basic Research Development Program of China (2016YFA0400503).
YAL thanks support by the European Research Council (ERC) under the European Union's Horizon 2020 research and innovation programme (682841 "ASTRUm").



\begin{thebibliography}{00}


\bibitem{Garcia73}J. D. Garcia, R. J. Fortner, and T. M. Kavanagh, Rev. Mod. Phys. {\bf 45} (1973) 111.
\bibitem{Shevelko18}V. P. Shevelko, Yu.  A. Litvinov, Th. St\"ohlker, and I. Yu. Tolstikhina, Nucl. Instrum. Methods B {\bf 421} (2018) 45.
\bibitem{Kaganovich05}I. D. Kaganovich, E.  A. Startsev, R. C. Davidson, S. R. Kecskemeti, A. Bin-Nun, D. Mueller, L. Grisham, R. L. Watson, V. Horvat, K. E. Zaharakis {\it et al.}, Nucl. Instrum. Methods A {\bf 544} (2005) 91.
\bibitem{Weber09}G. Weber, C. Omet, R. D. DuBois, O. de Lucio, Th. St\"ohlker, C. Brandau,  A. Gumberidze, S. Hagmann, S. Hess, C. Kozhuharov {\it et al.}, Phys. Rev. Accel. Beams {\bf 12} (2009) 084201.
\bibitem{DuBois07}R. D. DuBois, O. de Lucio, M. Thomason, G. Weber, Th. St\"ohlker, K. Beckert, P. Beller, F. Bosch, C. Brandau, A. Gumberidze {\it et al.}, Nucl. Instrum. Methods B {\bf 261} (2007) 230.
\bibitem{Shevelko11}V. P. Shevelko, I. L. Beigman, M. S. Litsarev, H. Tawara, I. Yu. Tolstikhina, and G. Weber, Nucl. Instrum. Methods B {\bf 269} (2011) 1455.
\bibitem{Shevelko01}V. P. Shevelko,  I. Yu. Tolstikhina, and Th. St\"ohlker, Nucl. Instrum. Methods B {\bf 184} (2001) 295.
\bibitem{Tolstikhina14}I. Yu. Tolstikhina, I. I. Tupitsyn, S. N. Andreev, and V. P. Shevelko, J. Exp. Theoret. Phys. {\bf 119} (2014) 1.
\bibitem{Henning08}W. F. Henning, Nucl. Phys. A {\bf 805} (2008) 502c.
\bibitem{Yang13}J. C. Yang, J. W. Xia, G. Q. Xiao, H. S. Xu, H. W. Zhao, X. H. Zhou, X. W. Ma, Y. He, L. Z. Ma, D. Q. Gao {\it et al.}, Nucl. Instrum. Methods B {\bf 317} (2013) 263.
\bibitem{Lapicki08}G. Lapicki, J. Phys. B: At. Mol. Opt. Phys. {\bf 41} (2008) 115201.
\bibitem{Shao17}C. Shao, D. Yu, X. Cai, X. Chen, K. Ma, J. Evslin, Y. L. Xue, W. Wang, Y. S. Kozhedub, R. C. Lu {\it et al.}, Phys. Rev. A {\bf 96} (2017) 012708.
\bibitem{Eichler07}J. Eichler and Th. St\"ohlker, Phys. Rep. {\bf 439} (2007) 1.
\bibitem{Glorius19}J. Glorius, C. Langer, Z. Slavkovsk\'a, L. Bott, C. Brandau, B. Br\"uckner, K. Blaum, X. Chen, S. Dababneh, T. Davinson {\it et al.}, Phys. Rev. Lett. {\bf 122}  (2019) 092701.
\bibitem{Herdrich17}M. O. Herdrich, G. Weber, A. Gumberidze, Z. W. Wu, and Th. St\"ohlker, Nucl. Instrum. Methods B {\bf 408} (2017) 294.
\bibitem{Mei15}B. Mei, T. Aumann, S. Bishop, K. Blaum, K. Boretzky, F. Bosch, C. Brandau, H. Bräuning, T. Davinson, I. Dillmann {\it et al.}, Phys. Rev. C {\bf 92} (2015) 035803.
\bibitem{Zamora17} J. C. Zamora, T. Aumann, S. Bagchi, S. B\"onig, M. Csatl\'os, I. Dillmann, C. Dimopoulou, P. Egelhof, V. Eremin, T. Furuno {\it et al.}, Phys. Rev. C {\bf 96},  (2017) 034617.
\bibitem{Schmid15} M. von Schmid, S. Bagchi, S. B\"onig, M. Csatl\'os, I. Dillmann, C. Dimopoulou, P. Egelhof, V. Eremin, T. Furuno, H. Geissel {\it et al.},  Phys. Scr. {\bf T166}, (2015) 014005.
\bibitem{Zamora16}J. C. Zamora, T. Aumann, S. Bagchi, S. Bönig, M. Csatlós, I. Dillmann, C. Dimopoulou, P. Egelhof, V. Eremin, T. Furuno {\it et al.}, Phys. Lett. B {\bf 763}, (2016) 16.
\bibitem{Yue19}K. Yue, J. T. Zhang, X. L. Tu, C. J. Shao, H. X. Li, P. Ma, B. Mei, X. C. Chen, Y. Y. Yang, X. Q. Liu {\it et al.}, Phys. Rev. C {\bf 100} (2019) 054609.

\bibitem{Koning03}A. J. Koning and J. P. Delaroche, Nucl. Phys. A {\bf 713}  (2003) 231.
\bibitem{Li08}X. Li and C. Cai, Nucl. Phys. A {\bf 801} (2008) 43.
\bibitem{Xu16}R. Xu, Z. Ma, Y. Zhang, and Y. Tian, Phys. Rev. C {\bf 94} (2016) 034606.
\bibitem{Xia02}J. W. Xia, W. L. Zhan, B. W. Wei, Y. J. Yuan, M. T. Song, W. Z. Zhang, X. D. Yang, P. Yuan, D. Q. Gao, H. W. Zhao {\it et al.}, Nucl. Instrum. Methods A {\bf 488}  (2002) 11.
\bibitem{CS}https://www.nndc.bnl.gov/exfor/
\bibitem{Thompson}FRESCO program, http://www.fresco.org.uk/
\bibitem{NRV}Nuclear Reactions Video Project, http://nrv.jinr.ru/nrv/
\bibitem{Doyle78}B. L. Doyle, U. Schiebel, J. R. Macdonald, and L. D. Ellsworth, Phys. Rev. A {\bf 17} (1978) 523.
\bibitem{Jonsson13}P. J\"onsson, G. Gaigalas, J. Biero\'n, C. Froese Fischer, and I. P. Grant, Comp. Phys. Comm. {\bf 184} (2013) 2197.
\bibitem{Klepper92}O. Klepper, F. Bosch, H. W. Daues, H. Eickhoff, B. Franczak, B. Franzke, H. Geissel, O. Gustafsson, M. Jung, W. Koenig {\it et al.}, Nucl. Instrum. Methods B {\bf 70} (1992) 427.
\bibitem{Litvinov11}Yu. A. Litvinov and F. Bosch, Rep. Prog. Phys. {\bf 74} (2011) 016301.
\bibitem{Najafi16}M. A. Najafi, I. Dillmann, F. Bosch, T. Faestermann, B. Gao, R. Gernh\"auser, C. Kozhuharov, S. A. Litvinov, Yu. A. Litvinov, L. Maier {\it et al.}, Nucl. Instrum. Methods A {\bf 836} (2016) 1.
\bibitem{Cipolla11}S. J. Cipolla, Comp. Phys. Comm. {\bf 182} (2011) 2439.



\bibitem{Sakaguchi17}H. Sakaguchi and J. Zenihiro, Prog. Part. Nucl. Phys. {\bf 97} (2017) 1.
\bibitem{Fattoyev18}F. J. Fattoyev, J. Piekarewicz, and C. J. Horowitz, Phys. Rev. Lett. {\bf 120} (2018) 172702.
\end{thebibliography}


\end{document}